\documentclass[aps,prb,twocolumn,groupedaddress,floatfix,showpacs]{revtex4}
\usepackage{graphicx}
\usepackage{bm}
\usepackage{color}
 
\newcommand{\Ang}{\AA$^{-1}$}
\newcommand{\half}{\ensuremath{^1\!\!/\!_2}}
\newcommand{\chem}[1]{\ensuremath{\mathrm{#1}}}
\newcommand{\GammaQ}{\ensuremath{\Gamma_{\bf q}}}
\newcommand{\ChiQ}{\ensuremath{\chi'_{\bf q}}}
\newcommand{\ChiO}{\ensuremath{\chi'_0}}
\newcommand{\GammaT}{\ensuremath{\Gamma(T)}}
\newcommand{\ChiT}{\ensuremath{\chi'(T)}}
\newcommand{\Sqw}{\ensuremath{S({\bf Q},\omega)}}
\newcommand{\Chiqw}{\ensuremath{\chi''({\bf q},\omega)}}
\def\ceptsi{{\rm CePt$_3$Si}}

\providecommand{\bfref}[3]{{\bf #1}, #3 (#2).} 

\def\jac{J. Alloys  Compd.\ }

\def\jpcm{J.\ Phys.:\ Condens.\ Matter }
\def\jmmm{J.\ Magn.\ Magn.\ Mater.\ }
\def\jpsj{J.\ Phys.\ Soc.\ Jpn.\ }

\def\pb{Physica B }
\def\prb{Phys.\ Rev.\ B }
\def\prl{Phys.\ Rev.\ Lett.\ }
\def\ssc{Solid State Commun.\ }
\def\zpb{Z. Phys. B }

\begin{document} 
\title{Low-energy magnetic response of the noncentrosymmetric heavy-fermion 
superconductor \ceptsi\ studied via inelastic neutron scattering}
\author{B. F\aa k}
\author{S. Raymond}
\author{D. Braithwaite}
\author{G. Lapertot}
\affiliation{Commissariat \`a l'Energie Atomique, INAC, SPSMS, 38054 Grenoble, France}
\author{J.-M. Mignot}
\affiliation{Laboratoire L\'eon Brillouin, CEA-CNRS, CEA-Saclay, 91191 Gif-sur-Yvette, France}
\date{19-August-2008; Revised: 2-Oct-2008}

\begin{abstract}
The low-energy magnetic excitations of the noncentrosymmetric heavy-fermion superconductor 
\ceptsi\ have been measured with inelastic neutron scattering on a single crystal. 
Kondo-type spin fluctuations with an anisotropic wave vector dependence are observed in the paramagnetic state. 
These fluctuations do not survive in the antiferromagnetically ordered state below $T_N=2.2$ K 
but are replaced by damped spin waves, 
whose dispersion is much stronger along the $c$-axis than in other directions.  
No change is observed in the excitation spectrum or the magnetic order 
as the system enters the superconducting state below $T_c\approx0.7$ K. 
\end{abstract}

\maketitle

\section{Introduction}
\label{SecIntro}
The discovery\cite{Bauer04} of unconventional superconductivity in the noncentrosymmetric heavy fermion compound \ceptsi\ 
has triggered a large amount of recent experimental  and theoretical activity 
(for corresponding reviews, see Refs.\ \onlinecite{Bauer07} and \onlinecite{Fujimoto07rev}, respectively). 
The main interest arises from the symmetry properties of the superconducting state. 
The lack of inversion symmetry leads to an antisymmetric spin-orbit coupling,
which mixes spin singlet and spin triplet Cooper-pairing channels.\cite{Gorkov01}
An outstanding question concerns the pairing mechanism of the superconductivity. 
Magnetism is an obvious candidate,\cite{Mathur98,Monthoux07}
since the Cooper pairs are formed by heavy quasiparticles.\cite{Bauer04}

We have therefore performed inelastic neutron scattering experiments 
to study the magnetic properties on a microscopic level of single crystalline \ceptsi. 
Our main findings are the following: 
Above the ordering temperature, 
quasielastic spin fluctuations with an anisotropic wave vector dependence are observed. 
In contrast to several other heavy-fermion systems, 
like \chem{CePd_2Si_2},\cite{vanDijk00}
\chem{CeIn_3},\cite{Knafo03}
and \chem{UPd_2Al_3},\cite{Bernhoeft98}
these spin fluctuations {\it do not} persist in the antiferromagnetically ordered state, 
where damped spin waves are the only low-energy excitations observed. 
No change is observed in the magnetic response as the system enters the superconducting state. 

\ceptsi\ crystallizes in the noncentrosymmetric tetragonal space group $P4mm$ (\#99) 
with room temperature lattice parameters $a=4.072$ and $c=5.442$ \AA. \cite{Tursina04}
Inversion symmetry is broken by the absence of a mirror plane perpendicular to the $c$-axis. 
The magnetic Ce atoms are in the 1(b) position at (\half,\half,0.1468) with $4mm$ point symmetry. 
Magnetic susceptibility measurements\cite{Bauer04}  give an effective moment (in the high-temperature limit) 
of $\mu_{\rm eff}=2.54$ $\mu_B$
and a Curie-Weiss constant $\theta_{\rm CW}$ of -45 K (-75 K) for a magnetic field along the $a$- ($c$-) axis,\cite{Takeuchi04}
indicating dominating antiferromagnetic interactions. 
The anisotropy at low temperatures is small, of the order of 15\%.\cite{Takeuchi04}
A relatively large mass enhancement is indicated by the linear term 
in the normal-state specific heat of 390 mJK$^{-2}$mole$^{-1}$,\cite{Bauer04}
while the Kondo temperature has been estimated to $T_K\approx 10$ K.\cite{Bauer04,Adroja05}

\begin{figure}[b]
\includegraphics[width=.5\columnwidth]{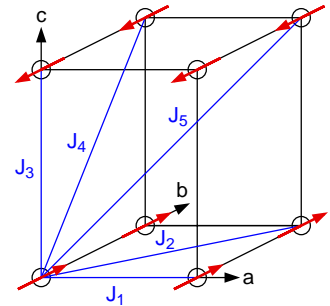} 
\caption{(Color online) Tetragonal unit cell of \ceptsi\ showing only the Ce atoms. 
The principal exchange integrals, $J_1$--$J_5$, are shown by blue lines
and the magnetic structure suggested by Ref.\ \onlinecite{Metoki04} by thick red arrows. 
The collinear moment direction can be anywhere in the $a$--$b$ plane.}
\label{FigCell}
\end{figure}

Long-range antiferromagnetic order sets in below the N\'eel temperature $T_N=2.2$ K.\cite{Bauer04}
Neutron diffraction measurements on single crystals show a magnetic propagation vector of ${\bf k}=(0,0,\half)$.\cite{Metoki04}
The corresponding magnetic structure consists of sheets of ferromagnetically ordered moments aligned in the $a$--$b$ plane 
with a moment of 0.16(1) $\mu_B$, which are stacked antiferromagnetically along the $c$-axis.\cite{Metoki04}
A likely magnetic structure is shown in Fig.\ \ref{FigCell}. 
Superconductivity occurs below $T_c=0.75$ K.\cite{Bauer04}
NMR\cite{Yogi04} and  $\mu$SR\cite{Amato05} measurements show that
superconductivity coexists on a microscopic scale with long-range homogenous bulk antiferromagnetic order.
The unconventional nature of the superconductivity is manifested in many measurements: 
The upper critical field, $H_{c2}$, is at least three times higher than the paramagnetic (Pauli) limiting field. 
On powder samples, $H_{c2}\!\approx\!3.3$--5 T,\cite{Bauer04,Izawa05,Higemoto06}
while measurements on single crystals give 2.7 (3.2) T for $H//a$ ($H//c$).\cite{Yasuda04,Takeuchi07}
The existence of line nodes in the superconducting gap has been inferred from measurements of
the magnetic penetration depth,\cite{Bonalde05}
specific heat,\cite{Takeuchi07,Bauer07}
and thermal transport.\cite{Izawa05}
Measurements of the spin-lattice  relaxation rate $1/T_1$ 
also indicate line nodes at low temperatures.
It is presently not clear whether $1/T_1$ displays a coherence peak\cite{Yogi04,Yogi06} or not.\cite{Ueda07}
The Knight shift remains constant on entering the superconducting state.\cite{Higemoto06,Ueda07}

There has been some discussion in the literature on the crystal field scheme. 
The $J=5/2$ multiplet of the stable $4f^1$ Ce$^{3+}$ ion splits in tetragonal point symmetry into three doublets. 
Neutron scattering measurements \cite{Adroja05,Metoki04} on powder samples 
show a broad magnetic excitation at about 16 meV and some intensity at or below 1 meV. 
If the broad excitation is interpreted as two peaks at 13 and 20 meV, 
the crystal-field level scheme is in agreement with entropy considerations.\cite{Adroja05} 
In this work, we will adopt the level scheme of Ref.\ \onlinecite{Adroja05}
which has a doublet ground state $a|\pm 5/2\rangle + b|\mp 3/2\rangle$ with $a=-0.59$ and $b=0.81$, 
a first excited state $b|\pm 5/2\rangle - a|\mp 3/2\rangle$ at 13 meV,
and a second excited state $|\pm 1/2\rangle$ at 20 meV.
The ground state doublet carries only a relatively small spin, 
which  implies that the ordered moment  is reduced due to crystal-field effects 
rather than Kondo screening. 

\begin{figure}
\includegraphics[width=.85\columnwidth]{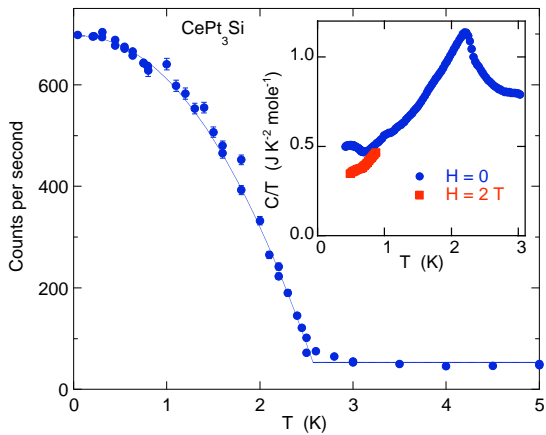} 
\caption{(Color online) Temperature dependence of the magnetic Bragg peak intensity at ${\bf Q}=(0,0,\half)$. 
The line is a guide to the eye. 
Critical magnetic scattering is observed between $T_N\!\approx\!2.5$ K and about 3 K.
The inset shows the specific heat divided by temperature at zero field (blue circles) and $H=2$ T (red squares) 
of the bottom part of the \ceptsi\ crystal used for the neutron scattering measurements. 
Onset of superconductivity occurs at $T_c=0.7$ K. }
\label{FigMofT}
\end{figure}

\section{Experimental}
\label{SecExp}
A cylindrical single crystal of diameter 6--7 mm, length 16 mm, and weight 7 g was grown in an image furnace. 
The crystal mosaic was 1.7$^\circ$. 
The sample was aligned with the $b$-axis (which was approximately parallel to the cylinder axis) 
vertical.
Specific heat measurements  on thin slices cut from the top and bottom of the single crystal used for the neutron scattering measurements 
show a pronounced  anomaly at $T_N=2.2$ K signaling the antiferromagnetic transition. 
The onset of bulk superconductivity occurs at $T=0.7$ K (see inset of Fig.\ \ref{FigMofT}).

\begin{table}
\centering
\caption{Spectrometer configurations giving final wave vector and 
energy resolution of the incoherent scattering at the elastic position as full width at half maximum. }
\vspace{1.0ex}
\begin{tabular}{llrr}
\hline \hline
\# & Spectro &  $k_f$ (\Ang)& $\Delta E$  ($\mu$eV) \\
 \hline
A & IN12 & 1.06 & 40  \\
B & IN12 &  1.15 &  64 \\
C & IN12 & 1.45 &  164 \\
D & IN14 & 1.05 & 40  \\
E & 4F1 & 1.55 & 258  \\
F & 4F1 & 1.20 & 68  \\
\hline\hline
\end{tabular}
\label{TableConfig}
\end{table}

Neutron scattering measurements were performed on the cold neutron triple-axis spectrometers 
IN12 and IN14 at the Institut Laue-Langevin and on 4F1 at the Laboratoire L\'eon Brillouin. 
Pyrolytic graphite (002) was used for the vertically curved monochromator and for the doubly focused analyzer. 
A liquid-nitrogen cooled Be filter was used to reduce higher-order contamination. 
Measurements were performed with a fixed final wave vector $k_f$ 
as given in Table  \ref{TableConfig}, where the corresponding energy resolution is also stated. 
Data were corrected for higher-order contamination in the beam monitor. 

Our neutron scattering measurements at zero energy transfer confirm the antiferromagnetic ordering vector ${\bf k}=(0,0,\half)$. 
The temperature dependence of the magnetic Bragg peak intensity is shown in Fig.\ \ref{FigMofT}.
No anomaly in the measured intensity is observed at the superconducting transition. 
Since the magnetic phase transition is second-order (see Fig.\ \ref{FigMofT}),
representation analysis can be used to classify possible magnetic structures. 
We find that symmetry-allowed magnetic structures are collinear with the moments either along the $c$-axis or (somewhere) in the $a$--$b$ plane. 
The presence of magnetic reflections along the $c$-axis, such as ${\bf Q}=(0,0,\half)$, 
shows that the moments are in the $a$--$b$ plane. 
However, due to the formation of so-called S-domains, 
it is not possible to determine the direction of the spins in the $a$-$b$ plane using neutron scattering, 
unless an unequal domain population can be induced. 
Assuming the same structure as proposed by Metoki\cite{Metoki04} and shown in Fig.\ \ref{FigCell}, 
we find from the integrated intensities of the four magnetic Bragg peaks in the $(h0l)$ scattering plane accessible in configuration B that the ordered moment is 0.17(5) $\mu_B$, 
in excellent agreement with Ref.\ \onlinecite{Metoki04}. 

Our {\it inelastic} neutron measurements show the presence of damped spin waves below $T_N$ 
and quasielastic spin fluctuations above $T_N$. 
We will discuss these low-energy excitations in the following two sections. 
We use the notation  
${\bf k}_i-{\bf k}_f={\bf Q}={\bm\tau}+{\bf k}+{\bf q}$,  
where {\bf Q} is the total wave-vector transfer,
${\bm\tau}$ a reciprocal lattice vector, {\bf k} the magnetic ordering vector,
and ${\bf q}=(h,k,l)$ is the wave vector with respect to the magnetic zone center.

\begin{figure}
\includegraphics[width=.8\columnwidth]{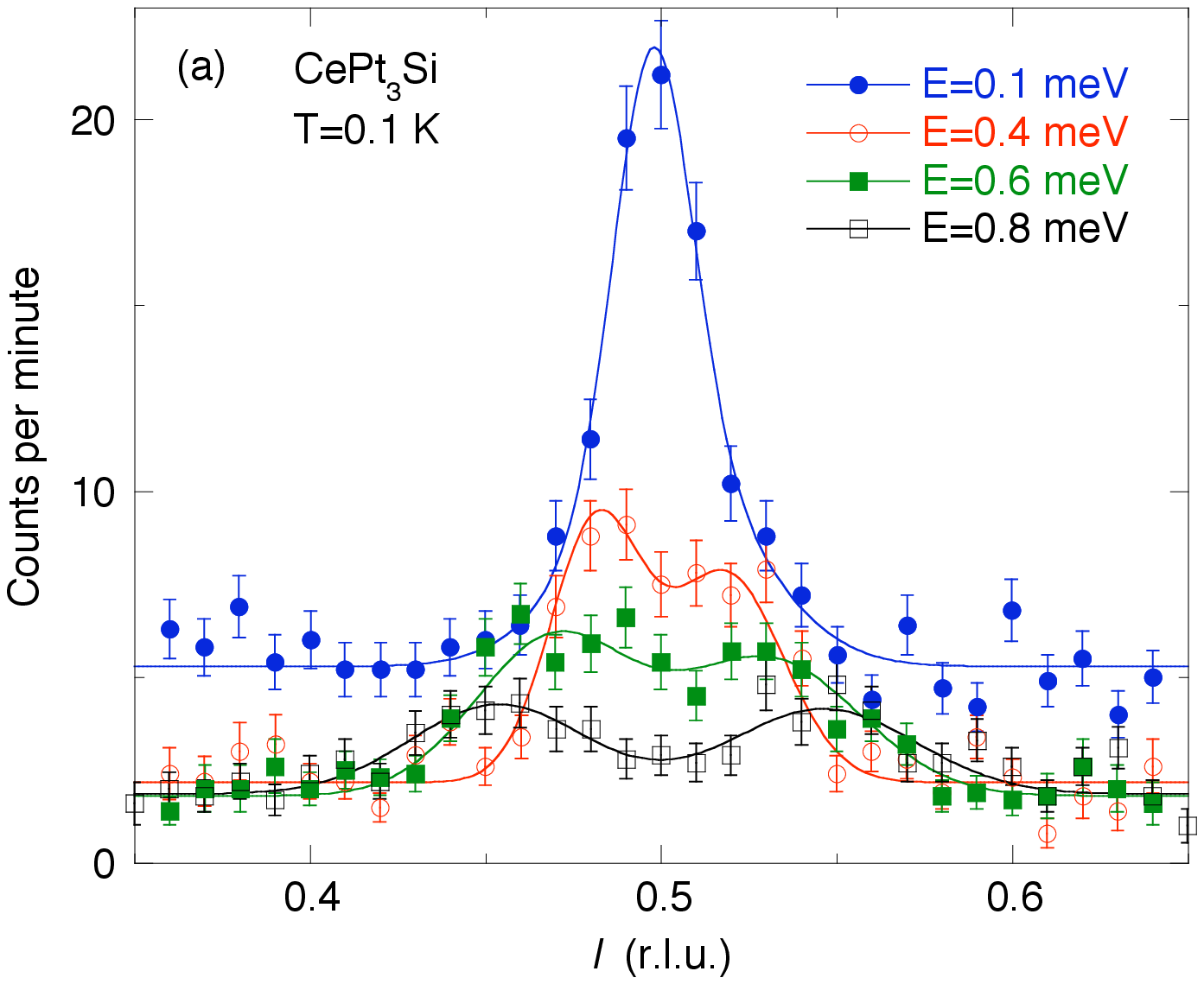}
\includegraphics[width=.8\columnwidth]{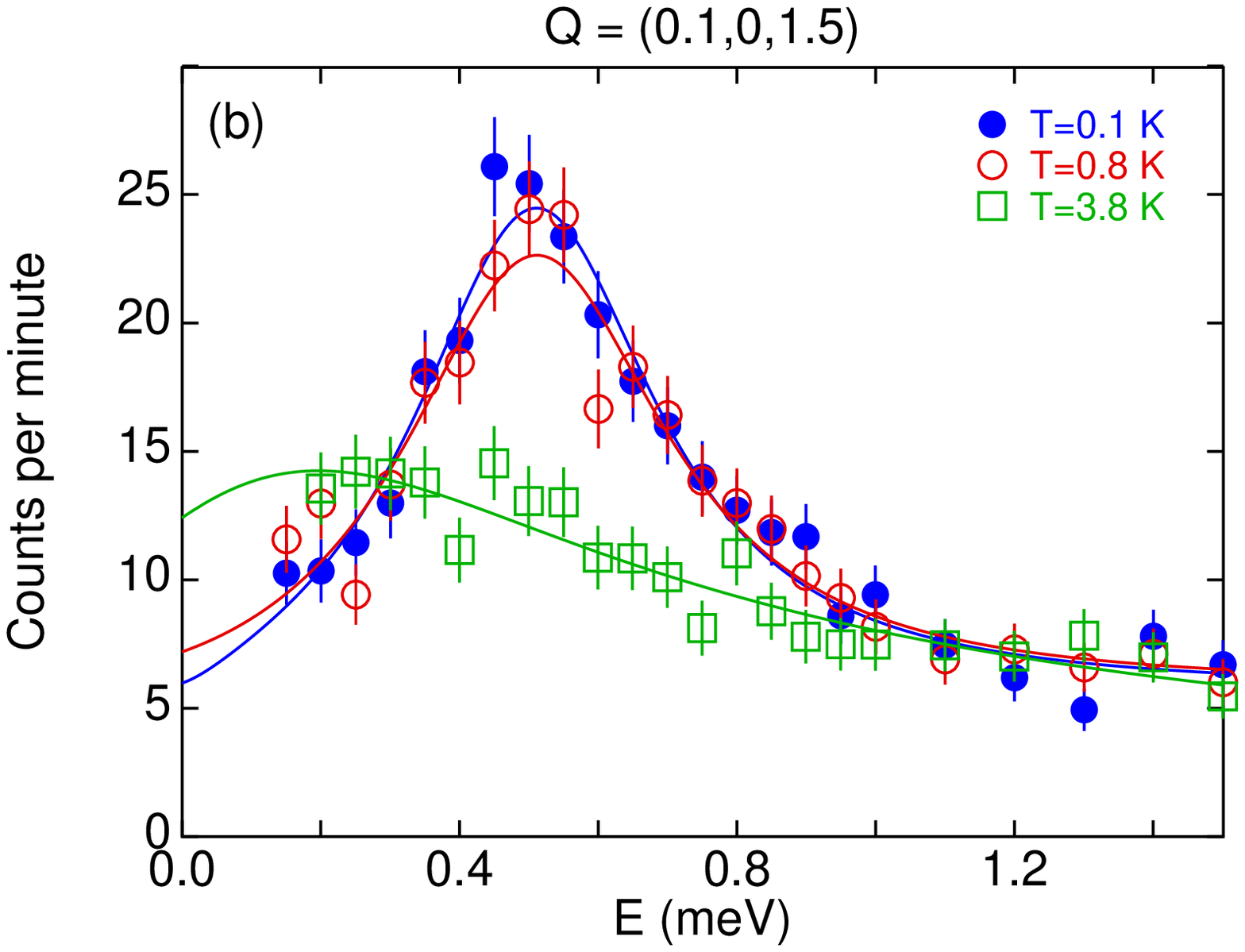}
\caption{(Color online) Spectra of spin-wave scattering. 
(a) $Q$-scans along $l$ measured in configuration D at ${\bf Q}=(0,0,l)$ 
for various energy transfers and fitted with two Gaussians. 
The higher background at the lowest energy transfer 
is due to the tail of the elastic incoherent scattering. 
(b) Energy scans measured in configuration B at ${\bf Q}=(0.1,0,1.5)$ 
at temperatures of 0.1, 0.8, and 3.8 K. 
The two lowest temperatures are fitted with a damped harmonic oscillator,
the highest with a quasielastic Lorentzian.}
\label{FigSWspectra}
\end{figure}

\begin{figure*}
\includegraphics[width=1.8\columnwidth]{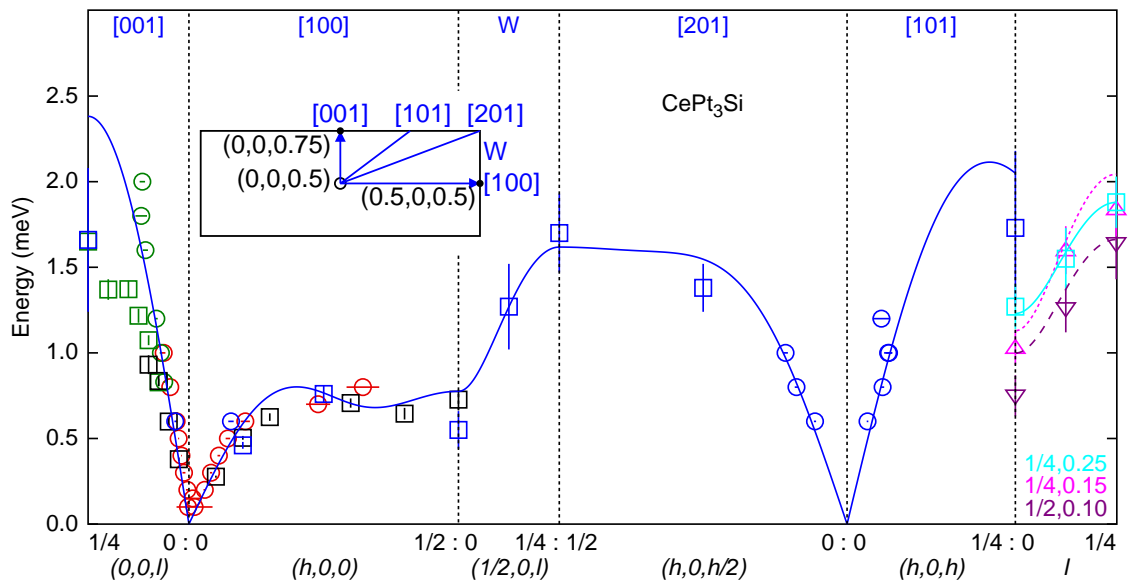}
\caption{(Color online) Spin wave dispersion along different {\bf Q} directions. 
Both $Q$ scans (circles) and energy scans (other symbols) were used. 
Different colors correspond to measurements performed on different instruments:
IN14 (red), 4F1 (green) and IN12 (black and blue);
see Table \ref{TableConfig}. 
The lines represent the fit of the spin-wave dispersion as discussed in the text. 
The right-hand-side panel shows the out-of-plane dispersion along $l$ for
${\bf Q}=(0.25,0.25,l)$, $(0.25,0.15,l)$, and $(0.5,0.10,l)$, respectively.
The inset shows the magnetic Brillouin zone in the $(h0l)$ plane 
and the main scan directions, also indicated at the top of the main figure.}
\label{FigDisp}
\end{figure*}

\section{Spin waves below $T_N$}
\label{SecSW}
In the antiferromagnetically ordered phase, damped spin waves are observed. 
Measurements were performed in configurations A-E of Table \ref{TableConfig}. 
The dispersion was measured along the main symmetry directions 
and additionally along a few off-symmetry directions, 
using a combination of energy and $Q$ scans.
Typical scans are shown in Fig.\ \ref{FigSWspectra}.
$Q$ scans were fitted by a sum of two Gaussian functions 
 while energy scans were fitted using a damped harmonic oscillator (DHO),\cite{DHO}
which was convoluted (in one dimension) with a Gaussian resolution function of width as specified in Table \ref{TableConfig}. 
The spin waves are damped already at the magnetic zone center and the damping increases with increasing $q$. 
The peak position of the DHO was used as spin-wave energy. 
The resulting spin-wave dispersion is shown in Fig.\ \ref{FigDisp}.
There is in general a good agreement between energy and $Q$ scans for the determination of the dispersion, 
although a small discrepancy can  be seen, in particular along the $l$-direction. 
This is probably due to the simplified treatment of the spin-wave damping. 
Measurements on different instruments are also in general good agreement. 
We find no evidence for a gap in the spin-wave dispersion. 
From $Q$ scans performed with high resolution (configuration D in Table \ref{TableConfig}),
the spin-wave gap is smaller than 0.1 meV, 
which is well below the gap of 0.23 meV inferred from specific-heat measurements.\cite{Adroja05}
However, the specific-heat measurements are stated to also agree with a gapless dispersion.\cite{Adroja05}

Most of the measurements of the spin waves were performed in the superconducting phase at $T<0.1$ K. 
Additional measurements at $T=0.8$ K, i.e. in the normal conducting and antiferromagnetically ordered state, 
did not reveal any additional broadening or change of the spin-wave energy 
(see e.g.\ Fig.\ \ref{FigSWspectra}). 
Measurements at even higher temperatures, $T=1.4$ K, give the same spin-wave energy as low-temperature measurements. 
Above the magnetic ordering temperature, 
no evidence of spin waves or any other relatively well-defined excitations was found in the energy range investigated ($E<5$ meV). 
Instead, broad quasielastic scattering of Kondo-type spin fluctuations were observed (see Fig.\ \ref{FigSWspectra}b), 
which are discussed in section \ref{SecSF}.

Since the splitting of the crystal-field levels ($\sim$10 meV)\cite{Adroja05}
is large compared to the spin-wave energies ($\sim$ 1 meV),
the dispersion of the latter can be calculated from the exchange-split crystal-field doublet ground state. 
In the absence of an observed gap in the spin-wave dispersion and in view of the collinear magnetic structure, we neglect single-ion anisotropy as well as anisotropic exchange, in which case 
 the dispersion relation for a two-sublattice commensurate antiferromagnet is\cite{Rossa}
\begin{equation}
\omega_q=  \sqrt{ [J({\bf k}) - J({\bf q})] [J({\bf k})-J({\bf k}+{\bf q})]},
\label{EqEsr}
\end{equation}
where $J({\bf q})= \sum_{\bf r}J({\bf r})\exp(-i{\bf q\cdot r})$ 
and the exchange integrals $J$ incorporate  the squared matrix element of the crystal-field ground state. 

The principal exchange integrals $J_1$--$J_5$ that enters Eq.~(\ref{EqEsr})  
are shown in Fig.\ \ref{FigCell}. 
In a simplest approach, only $J_1$ and $J_3$ are needed. 
Stability criteria for the magnetic structure with ${\bf k}=(0,0,\half)$ then require $J_3<0$ and $J_1>0$.  
In view of the stronger dispersion along the $c$-direction, it is clear that $|J_3|>|J_1|$. 
However, the steep dispersion along $[h00]$ near the magnetic zone center compared to the relatively small zone-boundary energy 
means that further exchange integrals in the basal plane are needed. 
In the spin-wave calculation, we have therefore included $J_6$ and $J_7$, 
taken to be along the $a$-direction corresponding to 2 and 3 units cells, respectively. 
Least-square fits of the spin-wave dispersion (\ref{EqEsr}) to the data, 
shown as lines in Fig.\ \ref{FigDisp}, leads to the following exchange integrals (in meV): 
$J_1=-0.13(3)$, 
$J_2=-0.04(2)$, 
$J_3=-0.71(5)$, 
$J_4=-0.12(1)$, 
$J_5=0.00(1)$, 
$J_6=0.012(3)$, and
$J_7=0.017(1)$. 
Since the exchange is governed by the oscillatory RKKY interaction, 
a large number of exchange integrals is not unexpected.
However, some comments are in place. 
To describe the dispersion along high-symmetry directions, 
the minimal set of exchange interactions  is 
$J_3<<0$, $J_4<0$, and $J_7>0$. 
The dispersion along the $[h00]$ direction is further improved by introducing $J_6>0$ in addition to $J_7$. 
Measurements along off-high-symmetry directions in the $(h0l)$ plane require in addition $J_1<0$ or $J_2<0$. 
We also performed measurements out of the $(h0l)$ plane by tilting the crystal 
(configuration C of Table \ref{TableConfig}). 
To describe the dispersion here, shown in the rightmost panel of Fig.~\ref{FigDisp},
both $J_1$ and $J_2$ are needed, with  $0>J_2>J_1$. 

While it would be tempting to compare the present exchange integrals with the oscillatory behavior of the RKKY interaction with distance, 
the complicated and anisotropic multi-band Fermi surface\cite{Hashimoto04} makes such a comparison meaningless.
Also, such calculations have not yet been confirmed experimentally, 
as de Haas van Alphen measurements on \ceptsi\ have not been able to locate 
the dominant Fermi surfaces.\cite{Hashimoto04,Takeuchi07}

The Curie-Weiss temperature can in a simplistic approach be expressed as
$\theta_{\rm CW}= \frac{2}{3}\sum_iz_iJ_i$,
where $z_i$ is the number of nearest neighbors for each bond $i$. 
Using the above values for $J_i$ (which include the classical spin-value $S$), 
we obtain $\theta_{\rm CW}=-23$ K, 
in fair agreement with  $-45$ K from bulk susceptibility measurements.\cite{Bauer04}

\section{Spin fluctuations above $T_N$}
\label{SecSF}
\begin{figure}
\includegraphics[width=.49\columnwidth]{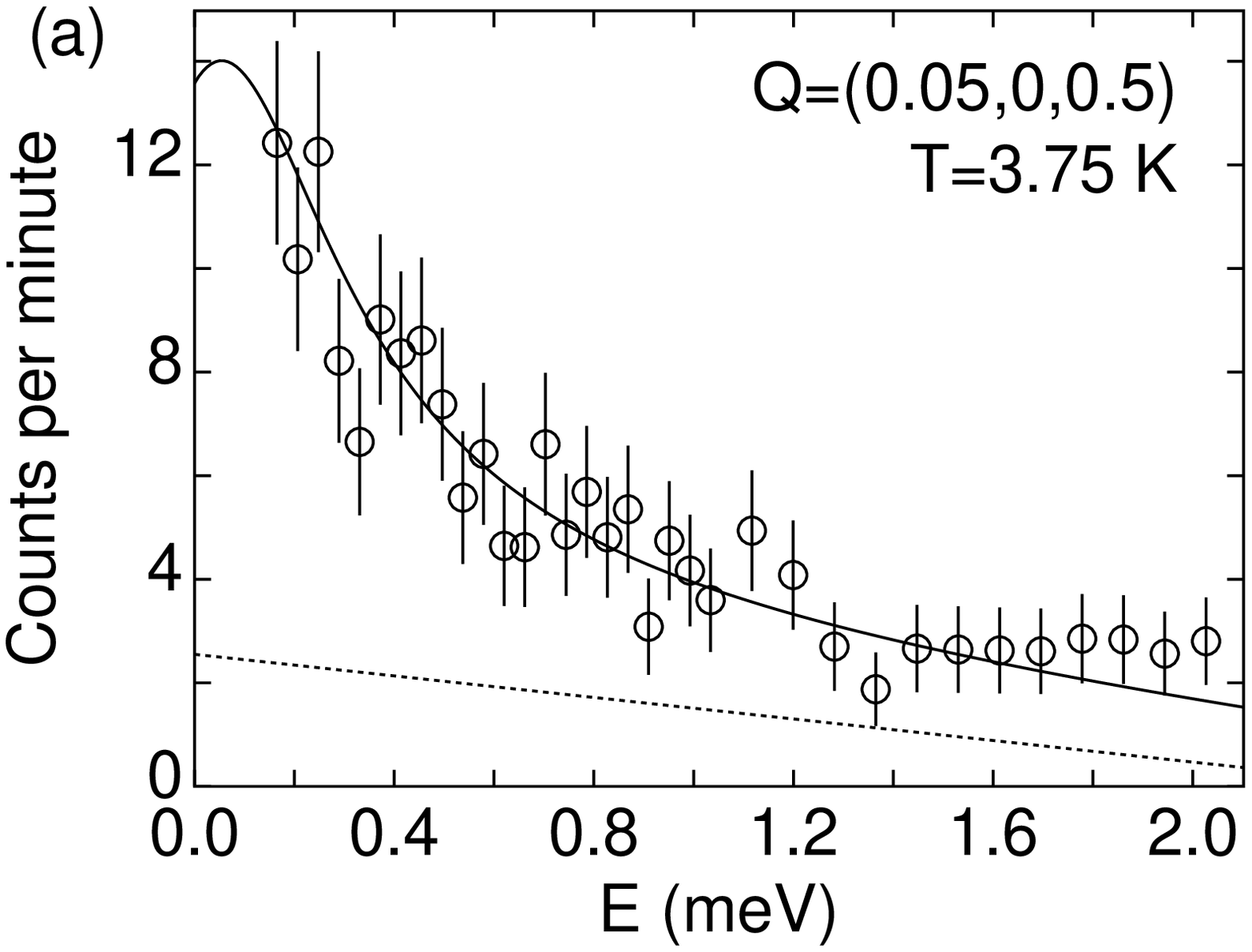} 
\includegraphics[width=.49\columnwidth]{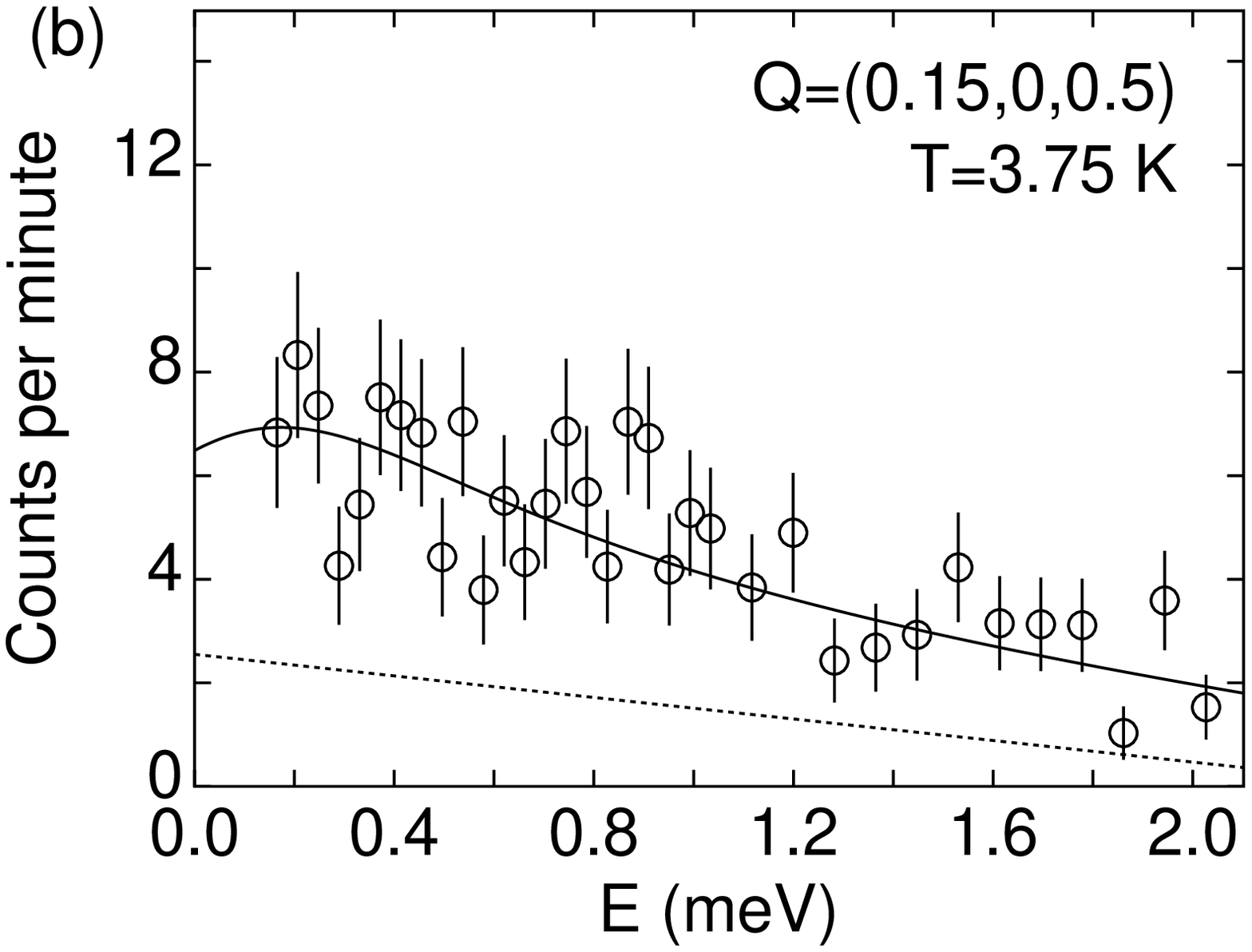} 
\includegraphics[width=.49\columnwidth]{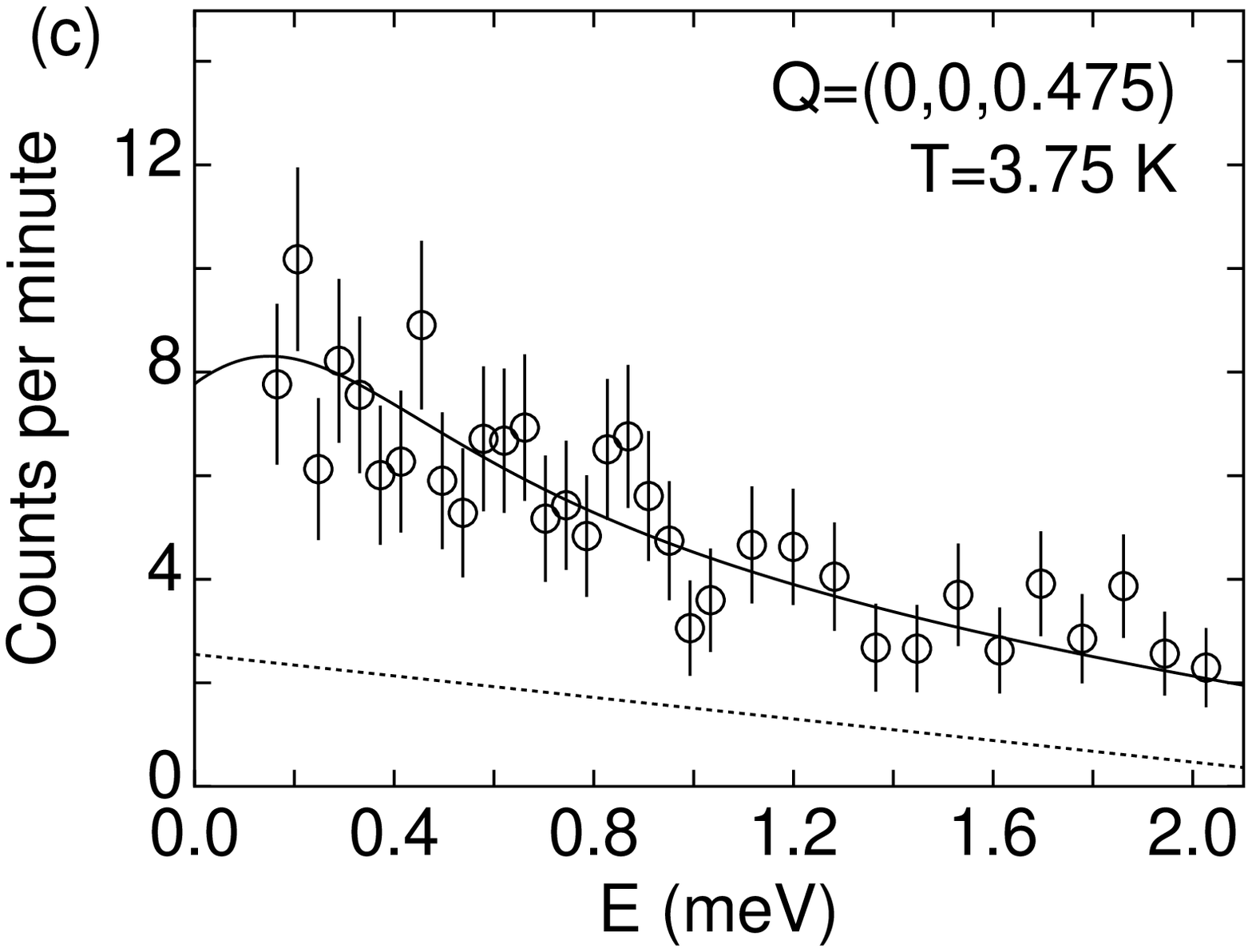} 
\includegraphics[width=.49\columnwidth]{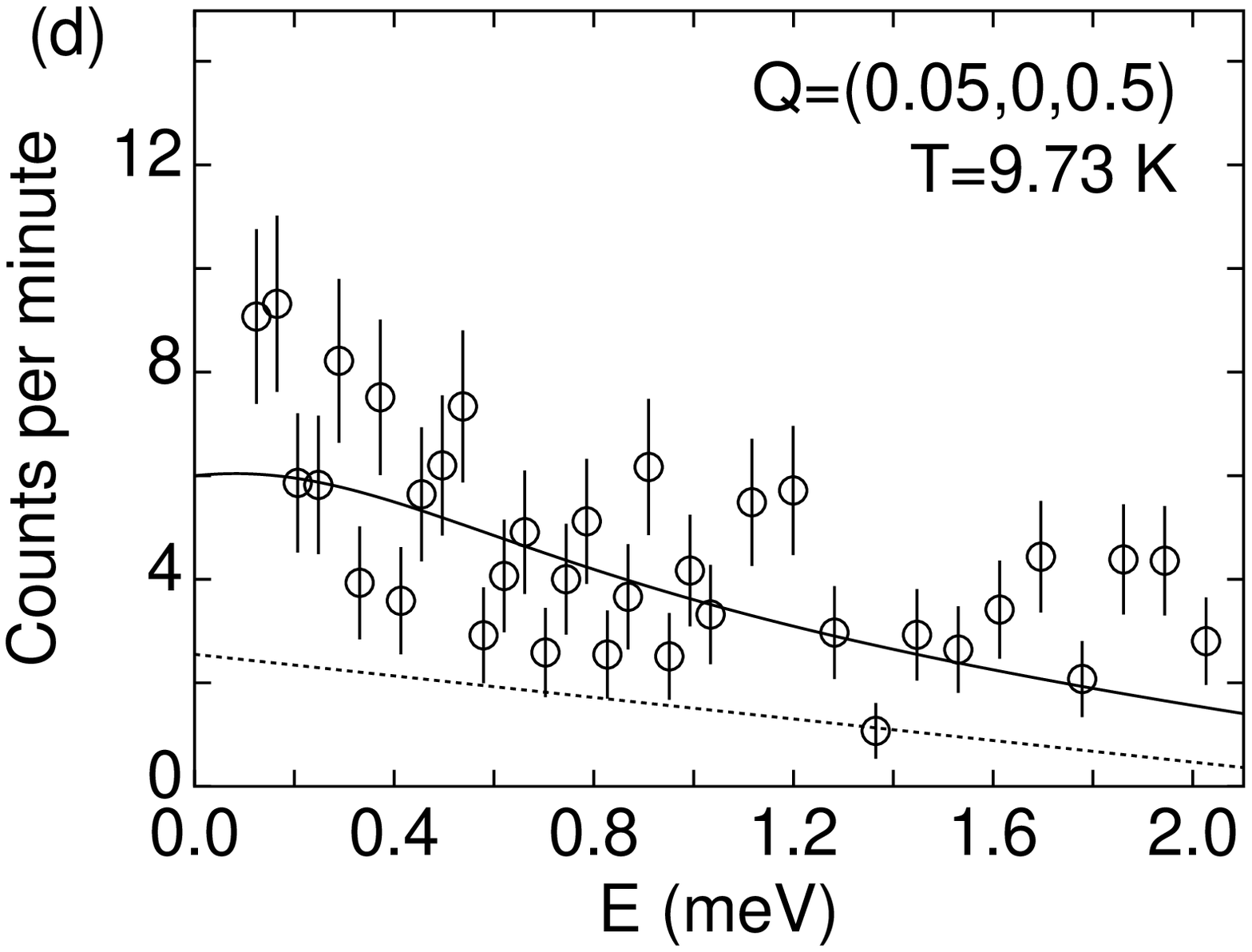} 
\caption{Spectra of spin-fluctuation scattering measured in configuration F. 
(a-c) $T=3.75$ K for different {\bf Q} values: 
near the magnetic zone center, along the $h$ direction, and along the $l$ direction. 
(d) Near the magnetic zone center at $T=9.73$ K. 
Solid lines are fits to a quasielastic Lorentzian, 
dotted line is the background level determined with the analyzer turned by 10$^\circ$.}
\label{FigSFspectra}
\end{figure}

Above $T_N$, the spin waves are replaced by spin fluctuations, 
which were measured in configuration F of Table \ref{TableConfig}. 
At temperatures just above $T_N$, 
critical magnetic scattering related to the thermal fluctuations driving the second-order phase transition was observed: 
this scattering decreased with increasing temperature and disappeared above $T=3$ K. 
At higher temperatures, quasielastic scattering related to Kondo-type spin fluctuations was observed. 
At $T=3.75$ K, we studied the {\bf q} dependence of the spin fluctuations in the vicinity of
the antiferromagnetic propagation vector ${\bf k}=(0,0,0.5)$. 
Typical scans are shown in Fig.\ \ref{FigSFspectra}. 
The data were well described by a quasielastic Lorentzian, 
\begin{equation}
\Chiqw =  
\frac {\omega \ChiQ \GammaQ} {\omega^2 + \GammaQ^2},
\label{EqLor}
\end{equation}
where the imaginary part of the dynamic susceptibility, \Chiqw,
is related to the observed neutron scattering intensity \Sqw\ via 
$\Chiqw =  [1-\exp(-\hbar\omega/k_BT)] \Sqw/f^2(Q)$, where $f(Q)$ is the magnetic form factor. 
The data were corrected for background, measured with the analyzer turned by 10$^\circ$, 
and the Lorentzian was convoluted with the instrumental resolution. 
 
\begin{figure}
\includegraphics[width=.85\columnwidth]{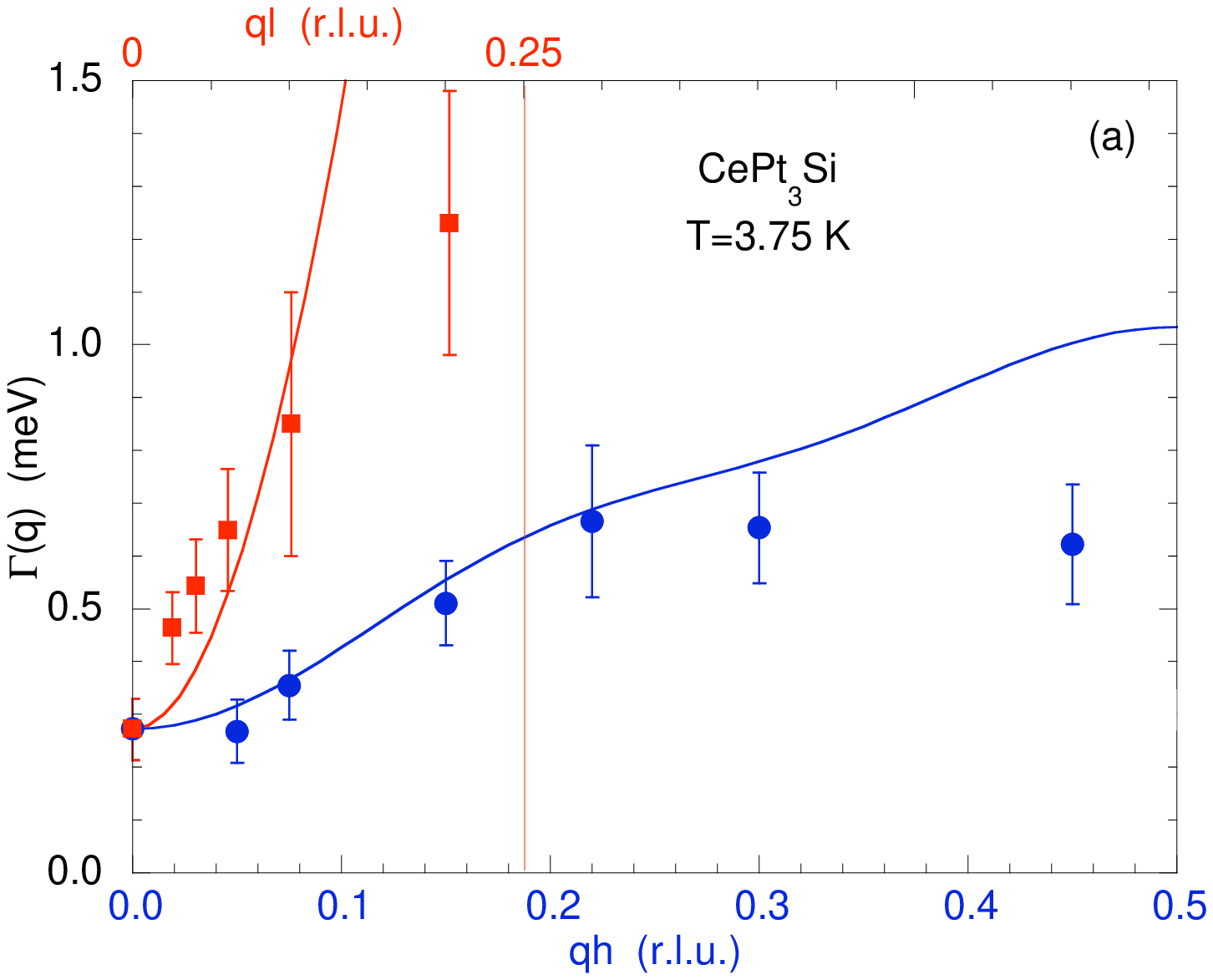}
\includegraphics[width=.85\columnwidth]{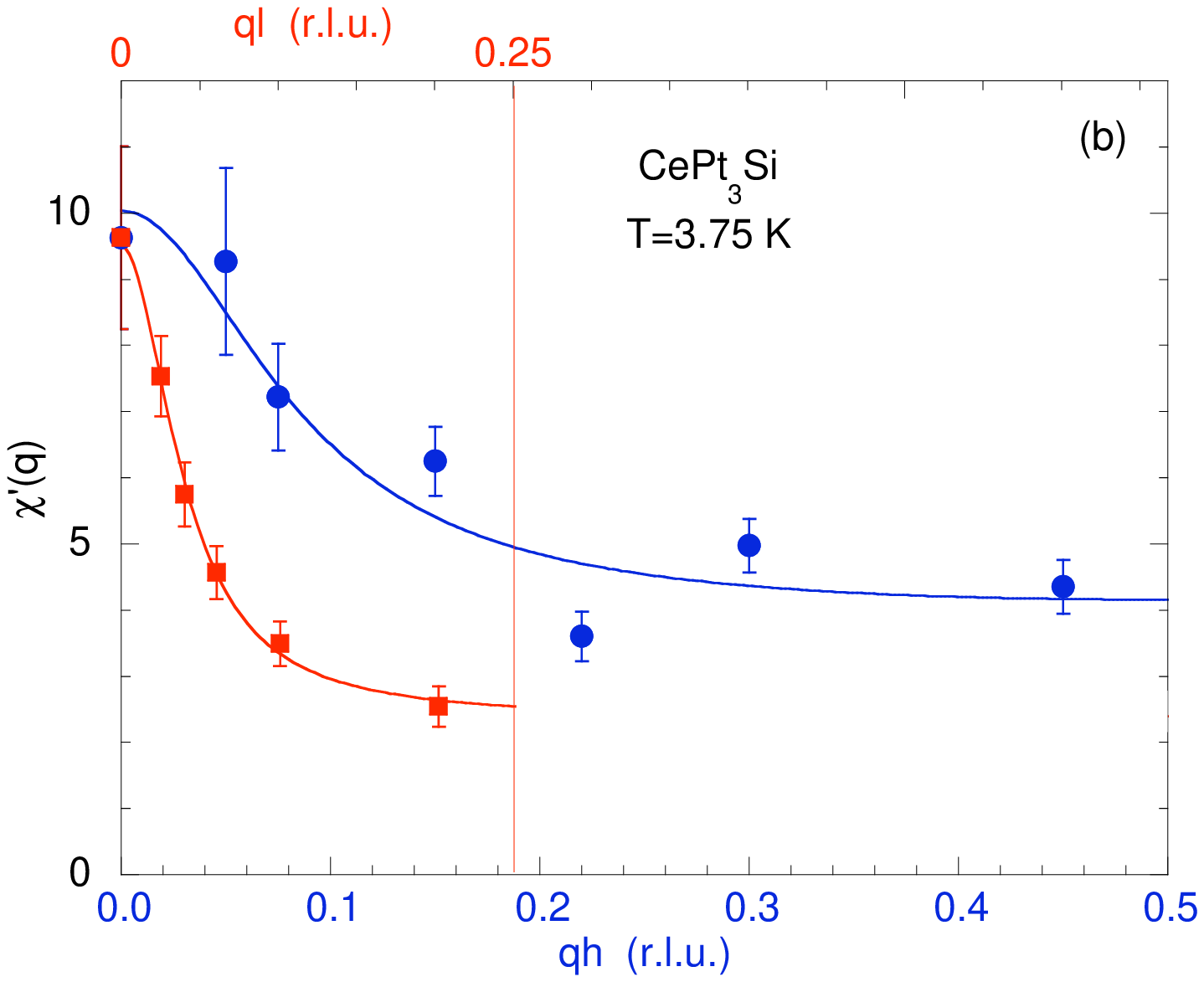}
\caption{(Color online) Wave vector dependence of 
(a) the quasielastic line width \GammaQ\ and 
(b) static susceptibility \ChiQ\ 
of the spin fluctuations at $T=3.75$ K.
Circles (squares) are for {\bf q} along the $h$ ($l$) direction, shown on the lower (upper) abscissa, respectively. 
The lines in (a) are \GammaQ\  estimated from the exchange integrals via Eq.\ (\ref{EqGammaVsJq})
and those in (b) are fits to Eq.\ (\ref{EqChiQ}).}
\label{FigGQ}
\end{figure}

The wave vector dependence of the fitted line width \GammaQ\ and the static susceptibility 
\ChiQ\ (after correction for the magnetic form factor) of Eq.\ (\ref{EqLor}) are shown in Fig.\ \ref{FigGQ}. 
The spin fluctuations are more strongly correlated along the $c$-axis than in the basal plane,
as seen from the rapid decay of \ChiQ\ along $l$. 
The data were fitted to (see Fig.\ \ref{FigGQ}b)
\begin{equation}
\ChiQ
= \frac {\ChiO} {1+(q_\alpha/\kappa_\alpha)^2} + \bar{\chi},
\label{EqChiQ}
\end{equation}
where the first term  is the standard Fermi-liquid susceptibility ($\alpha$ is $h$ or $l$)
and the second term is a constant, possibly related to interband contributions. 
We find the inverse correlation length 
$\kappa_h=0.09(3)$ and $\kappa_l=0.040(8)$ r.l.u.\ along the $a$- and $c$-axes, respectively,
which corresponds to an anisotropy in $Q$ space of a factor of 2. 
The energy scale of the quasielastic spin fluctuations, given by \GammaQ, 
increases with $q$, and with a much higher rate along the $c$-axis.
We find that the product \ChiQ\GammaQ\ is constant, 
in agreement with other heavy-fermion compounds as well as with theory.\cite{Kuramoto87}
We note that the $q$-averaged \GammaQ, which is of the order of 1 meV at $T=3.75$ K, 
leads to a linear contribution to the specific heat of a few hundreds of mJK$^{-2}$mole$^{-1}$,
in qualitative agreement with the measured value of 390 mJK$^{-2}$mole$^{-1}$.  

In a simple approach, it is possible to correlate the quasielastic line width \GammaQ\ of the low-energy Kondo spin fluctuations 
with the exchange integral $J({\bf q})$. 
Kuramoto\cite{Kuramoto87} has shown that the product of the  line width and the susceptibility is constant, 
i.e.\ $\ChiQ\GammaQ =  \ChiO\Gamma_0$, 
where the subscript 0 denotes the non-interacting value (where $J({\bf q})=0$). 
Combining this with the static susceptibility obtained in the random-phase approximation (RPA),
\begin{equation}
\ChiQ= \frac{\ChiO}{1-J_{\bf q}\ChiO},
\label{EqRPA}
\end{equation}
one finds that 
\begin{equation}
\Gamma_{\bf q}= \Gamma_0 - \ChiO\Gamma_0J_{\bf q}. 
\label{EqGammaVsJq}
\end{equation}
We take the product $\ChiO\Gamma_0$ from the average over {\bf q} of  $\ChiQ\GammaQ$ 
and $\Gamma_0$ from the measured value of \GammaQ\ at the antiferromagnetic zone center via Eq.\ (\ref{EqGammaVsJq}) 
with $J_{{\bf q}={\bf k}}$ from the above spin-wave calculation.
\GammaQ\ can then be calculated from Eq.\ (\ref{EqGammaVsJq}), 
which gives the solid lines in Fig.\ \ref{FigGQ}a, which are in remarkable good agreement with the measurements. 

\begin{figure}
\includegraphics[width=.85\columnwidth]{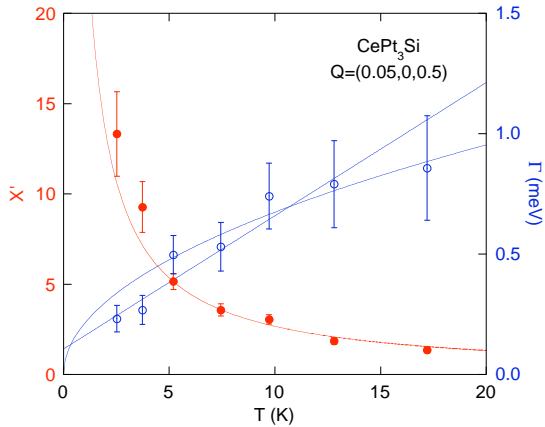}
\caption{(Color online) Temperature dependence of the fitted line width \GammaT\ (open circles) 
and static susceptibility \ChiT\ (closed circles) at ${\bf Q}=(0.05,0,0.5)$ for $T>T_N$.}
\label{FigTdep}
\end{figure}

The temperature dependence of the spin-fluctuation scattering above $T_N$ 
was measured close to the magnetic propagation vector, at ${\bf Q}=(0.05,0,0.5)$.
Typical scans are shown in Figs.\ \ref{FigSFspectra}a and d. 
The energy scale of the quasielastic spin fluctuations, given by \GammaQ, 
increases with temperature, as in other heavy-fermion systems. 
Figure \ref{FigTdep} shows that the line width is equally well
described by a linear temperature dependence
$\Gamma(T) = \Gamma(0) + cT$ with $\Gamma(0)=0.11$ meV
or a square root behavior $\Gamma(T) = c'\sqrt{T}$ 
while \ChiT\ follows approximately a $1/T$ behavior. 
The product \ChiT\GammaT\ has a slight tendency to decrease with increasing temperature. 

\section{Discussion}
\label{SecDisc}
The dynamic magnetic response of heavy-fermion systems at low temperatures 
is characterized by Kondo-type quasielastic spin fluctuations 
with a characteristic energy scale \GammaQ(T), 
which increases with increasing temperature 
and is minimal at wave vectors {\bf q} where the exchange $J({\bf q})$ is maximal, 
i.e.\ at wave vectors corresponding to incipient magnetic order (see e.g.\ Ref.\ \onlinecite{Kadowaki04}).
Surprisingly, it has been found in several heavy fermions that the spin fluctuations persist into the magnetically ordered phase, 
where they coexist with collective spin waves. 
Examples are  \chem{CePd_2Si_2},\cite{vanDijk00} \chem{CeIn_3},\cite{Knafo03}
and  \chem{UPd_2Al_3}.\cite{Bernhoeft98}
The two Ce compounds have quite similar properties to \ceptsi,
with a first excited crystal field level at 12--19  meV, an exchange of $\sim$0.4 meV, and an ordering temperature of $\sim$10 K. 
It is therefore somewhat surprising that in \ceptsi,
the spin fluctuations {\it do not} persist into the antiferromagnetically ordered phase.
The low-energy magnetic response of \ceptsi\ is thus characterized by damped spin waves below $T_N$ and quasielastic spin fluctuations above. 

A rather large number of heavy-fermion superconductors (HFSC) have been studied since the initial discovery of \chem{CeCu_2Si_2}.\cite{Steglich79}
A crucial question is whether there is a microscopic connection between the superconductivity and the magnetism. 
Many of the promising candidates to reveal such a coupling are superconducting only under pressure, 
which has seriously hampered their study with inelastic neutron scattering. 
We will therefore only discuss the rather limited number of heavy fermions that are superconducting at ambient pressure. 

In the archetypal HFSC \chem{CeCu_2Si_2}, 
it has been shown that magnetic order and superconductivity are mutually exclusive on a microscopic scale \cite{Stockert06muon} 
and that a spatial separation might occur.  
Two separate features appears to be characteristic for the superconducting state:
(i)~Short-range  elastic correlations close to ${\bf k}_{\rm afm}$
and (ii)~a gap of $\hbar\omega_0=0.2$ meV  in the quasielastic spin-fluctuation spectrum.\cite{Stockert08}
Both of these features disappear above the superconducting transition temperature of $T_c=0.6$ K.
The ratio of the gap energy to the transition temperature is  $\hbar\omega_0/(k_BT_c)\approx 4$.

\chem{CeCoIn_5} is a heavy-fermion superconductor with $T_c=2.3$ K and no long-range magnetic order.\cite{Petrovic01}
Inelastic neutron scattering measurements have shown the appearance of a sharp resonance peak 
at an energy of $\hbar\omega_0\!=$0.6 meV in the superconducting state, 
with a correlation length of about 10 \AA.\cite{Stock08}
The resonance is replaced by quasielastic spin fluctuations with a characteristic energy of $\Gamma\approx 1$ meV in the normal state. 
The value of $\hbar\omega_0/(k_BT_c)$ is of the order of 3. 

\chem{UPd_2Al_3} is a heavy fermion ($\gamma=210$ mJK$^{-2}$mole$^{-1}$) 
where superconductivity ($T_c\approx 1.9$ K) coexist with antiferromagnetic  order ($T_N=14.3$ K).\cite{Geibel91}
Below $T_N$, spin-waves coexist with quasielastic spin fluctuations. 
The latter develops a gap of 0.36 meV in the superconducting state;\cite{Bernhoeft98,Metoki98,Hiess06} $\hbar\omega_0/(k_BT_c)\approx 2.2$. 
In other uranium based heavy-fermion superconductors, such as 
\chem{URu_2Si_2},\cite{Broholm87}
\chem{UPt_3},\cite{Aeppli88}
\chem{UNi_2Al_3},\cite{Aso00}
and \chem{UBe_{13}},\cite{Coad00}
no changes in the magnetic excitation spectrum have been observed at $T_c$. 

Assuming that the ratio $\hbar\omega_0/(k_BT_c)$ is in the range of values 2.2--4 observed in other HFSC (see above), 
one would expect an excitation gap (or a resonance) of the order of 0.2 meV in \ceptsi.
As shown in Fig.\ \ref{FigSWspectra}, 
there is no indication of a gap of this size opening in the spin-wave dispersion.  
However, in the three examples discussed above, where the magnetic excitation spectrum is modified at $T_c$, 
it is the low-energy quasielastic response which is subject to a transfer of spectral weight. 
The absence of a significant quasielastic signal below $T_N$ may actually preclude the observation of such an effect in \ceptsi. 
Theoretical work show that the dynamic susceptibility in the superconducting state 
is determined by both the symmetry of the order parameter and the geometry of the Fermi surface.\cite{Joynt88,Lu92}
In this context, the mixed $s$- and $p$-wave superconductivity of \ceptsi\ appears as a much more complicated case than 
e.g.\  \chem{CeCoIn_5} ($d$-wave) and \chem{UPd_2Al_3} (AF $s$-wave) and would require further theoretical attention.

\section{Conclusion}
We have shown by inelastic neutron scattering measurements on single crystalline \ceptsi\ 
that the magnetically ordered state is characterized by damped spin waves with quite strongly anisotropic exchange interactions. 
In contrast to some other heavy-fermion systems, there is no evidence of spin fluctuations in the ordered state. 
Above $T_N$, Kondo-type spin fluctuations are observed, 
with an anisotropy in $Q$ that reflects the anisotropy in the exchange integrals, 
assuming a simple RPA scenario.  
We have not observed any change in the magnetic order  or in the magnetic excitations (in terms of a gap or a resonance) 
as \ceptsi\ enters the superconducting state. 
This may be related to the absence of quasielastic scattering in the magnetically ordered phase.

\begin{acknowledgments}
We thank K.~Mony for invaluable assistance in sample preparation and alignment, 
B.~Vettard and Ph.~Boutrouille for cryogenic and technical support, 
and P.~C.~Canfield, M.~Enderle, F.~Givord, V.~P.~Mineev,  J.~Schweizer, 
and M.~E.~Zhitomirsky for helpful discussions. 
Part of the sample preparation was made at the Ames Laboratory, 
which was supported by the Department of Energy, Basic Energy Sciences under Contract 
No.\ DE-AC02-07CH11358,
while part of the neutron scattering measurements were performed at the Institut Laue-Langevin, Grenoble, France. 
\end{acknowledgments}

\end{document}